# Survival Probability of an Excited State in the Bixon-Jortner Model

James P. Lavine
Department of Physics
Georgetown University
37th and O Sts. NW
Washington, DC 20057 USA

**ABSTRACT**

When the initial state of a quantum mechanical system is an excited state, then it is expected that the occupation, or survival, probability of that state will decrease. This is studied numerically within the Bixon-Jortner model, which was introduced to model intramolecular radiationless transitions. Here a finite set of states is used and for a fixed number of states, the parameters of the model are the energy level separation and the strength of the transition matrix element. All three of these are varied to see their effects on the survival probability. After a short interval of time, the survival probability decay is often found to be an exponential. But the survival probability is then found to increase with further time and then decrease in a pattern that continues in time. This repopulation is a general feature when a countable set of states is present.

## I  INTRODUCTION

If a system is prepared in an initial state, then how does the occupation or survival probability, $p_s(t)$, evolve with $t$ the time? For example, is it exponential in time? The Wigner-Weisskopf method [1] was an early approach to this problem. Their model and its exponential time dependence are explained in Appendix C of Le Bellac[2] and in Section 5.3 of Barnett and Radmore.[3] Yet investigations have established that the decays at short times and at large times depart from an exponential.[4,5] A further concern is whether the initial state becomes repopulated,[4-7] which is often referred to as recurrence or regeneration. In sum, these concerns argue against pure exponential decay and suggest that exploring model systems may supply insights into $p_s(t)$.

The present work investigates the Bixon-Jortner model[8], which was introduced to model intramolecular radiationless transitions. This model treats the decay of an initial state into a set of other states in a highly-idealized manner. Here the set of other states is assumed to be finite. Scherer[9] provides a short introduction and illustration of the model. The initial state is connected by a transition matrix element to each of the other states, but these other states are not connected to each other. Previous work with the Bixon-Jortner model used a range of solution methods.[3,8-14] References 8 to 12 find the energy levels of the Hamiltonian that includes the transition matrix elements. These lead to the eigenvectors and equations for the time-dependence of the wave functions. Barnett and Radmore[3] apply Laplace transforms and Mello, Pereyra, and Kumar[13] use a random matrix method. In addition, Scherer and Fischer[14] develop the propagator for the Bixon-Jortner model.

This work treats the time development of the occupation probabilities of the states as an initial value problem in time. Section II turns Schrödinger's Equation for the Bixon-Jortner model into a finite set of coupled, linear, first-order, ordinary differential equations, which are solved numerically. Ideally, the initial state should decay into a continuum of states, but only a finite number is treated here. Section III shows that 25 states suffice to approximate a continuum. This section has additional numerical results for the behavior of the survival probability $p_s(t)$. Even the cases that do exhibit an exponential decay in $p_s(t)$ for a range of times, then show a $p_s(t)$ that repeatedly grows and decreases as the time $t$ increases. This latter behavior is a direct result of treating only a countable number of states. Finally, Section IV discusses the results and suggests some future avenues to explore.

## II THE BIXON-JORTNER MODEL AND THE NUMERICS

The model[8] starts with a state denoted by the subscript *s* with a population of 1 and with all the other states unoccupied. State *s* decays into a set of states that are labelled by $k = \{-m, -m+1, \ldots, -1, 0, 1, \ldots, m-1, m\}$. Here, *m* is usually set to 12, so state *s* decays into 25 other states. Please note the state *s* is not one of the *k*-states. The total number of states is $n = 2m + 2 = 26$.

The matrix element for a transition from state *s* to any state *k* is

$$\bar{V} = \langle k|\hat{V}|s\rangle/\hbar = \langle s|\hat{V}|k\rangle/\hbar, \tag{1}$$

and the matrix element is real and a constant that is the same for all *k*. The hat indicates an operator. There are no transitions between any of the *k*-states. Next, we need a Hamiltonian. Let

$$\hat{H} = \hat{H_0} + \hat{V}, \tag{2}$$

and

$$\hat{H_0}|i\rangle = E_i|i\rangle. \tag{3}$$

Here *i* is *s* or a *k*. These states are assumed to be orthonormal and the Hamiltonian is taken to be independent of time. This summarizes the Bixon-Jortner model.

The model is deceptively simple-looking. Solutions for a finite number of states require numerical methods and the present work turns Schrödinger's Equation into a set of coupled, linear, first-order, ordinary differential equations. This approach is often used for time-dependent Quantum Mechanical problems[15] and is referred to here as ODE for ordinary differential equations.

The ODE method starts by writing the time-dependent solution to Schrödinger's Equation in terms of the eigenkets of $\hat{H_0}$

$$|\Psi(t)\rangle = \sum_i x_i(t)|i\rangle. \tag{4}$$

Now applying the time-dependent Schrödinger's Equation leads to

$$i\hbar \frac{\partial}{\partial t} \sum_i x_i(t)|i\rangle = (\hat{H_0} + \hat{V}) \sum_i x_i(t)|i\rangle. \tag{5}$$

Next multiple from the left with $\langle j|$ and use the orthonormality to find

$$i\hbar \dot{x}_j(t) = \hbar \omega_j x_j + \sum_i x_i(t)\langle j|\hat{V}|i\rangle \,, \tag{6}$$

where $E_j = \hbar \omega_j$. Now suppose $j \neq s$, then

$$i\hbar \dot{x}_j(t) = \hbar \omega_j x_j + x_s(t)\langle j|\hat{V}|s\rangle \,. \tag{7}$$

When $j = s$, then

$$i\hbar \dot{x}_s(t) = \hbar \omega_s x_s + \sum_i x_i(t)\langle s|\hat{V}|i\rangle \,. \tag{8}$$

The division of both sides by $\hbar$, explains the definition of $\bar{V}$ in Eq. (1). To be specific,

$$\omega_k = \omega_s + k\varepsilon \,. \tag{9}$$

Here $\omega_s$ is set equal to 0 and the successive $k$ energy levels are separated by the parameter $\varepsilon$.

Hence, 26 coupled equations result from Eqs. (7) and (8). These are solved numerically by the routine NDSolve of Mathematica.[16] The maximum time step used in NDSolve is 0.001. As a check on the numerics, the total occupation probability

$$p_{tot}(t) = \sum_i x_i^*(t) x_i(t) \,, \tag{10}$$

is tracked and it stays well within $1.000000 \pm 0.0000025$. Thus, probability is conserved by the numerical method. In Eq. (10), $*$ indicates the complex conjugate. Examples of the numerical solutions are presented in Section III.

## III NUMERICAL RESULTS

The ODE method treats the initial state $s$ and the 25 $k$-states. The ODE method permits viewing the occupation probabilities for each of the $k$-states and these are referred to as $p_k(t)$. This helps probe how the regeneration or repopulation occurs. The first two cases are motivated by Figs. 2.4 and 2.5 of Barnett and Radmore[3]. Figure 1 has the survival probability for state $s$, $p_s(t)$, for $t = 0$ to 60 when $\bar{V} = 0.10$ and $\varepsilon = 0.25$. After a very short time interval, $p_s(t)$ actually drops exponentially in time. This is verified by the semi-logarithmic plot of Fig. 2 for $t = 0$ to 20, which also displays the departure from the exponential when $t \to 0$. Figure 1 also shows that the occupation probability of the $k = 0$ and $k = 1$ states varies strongly with $p_s(t)$.

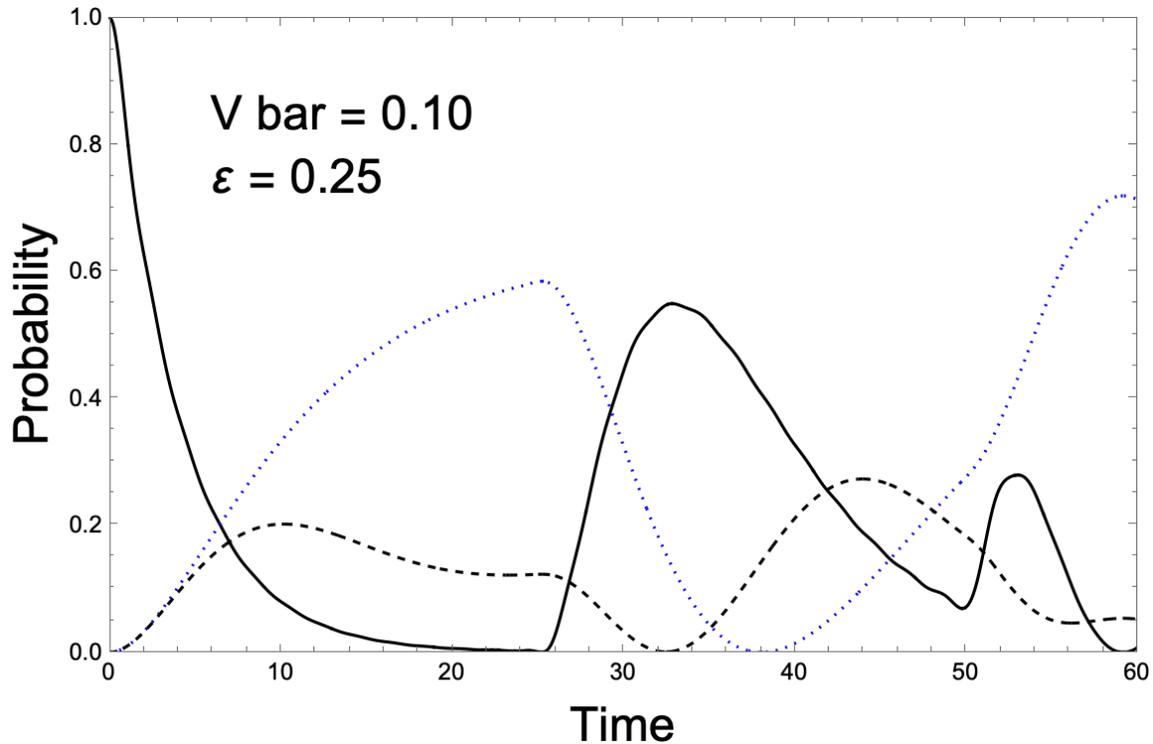

Fig. 1. The occupation of the initial state versus time for $\bar{V} = 0.10$ and $\varepsilon = 0.25$ with $n = 26$. The solid line is $p_s(t)$, the dotted line (blue) is the occupation probability for state $k = 0$, and the dashed line is the occupation probability for $k = 1$.

The second case uses $\bar{V} = 0.04$ and $\varepsilon = 0.25$ and Fig.3 shows the results are similar to those in Fig. 1, although the initial drop in $p_s(t)$ is much less. This drop is also exponential and the interplay between state $s$ and the $k = 0$ state is quite pronounced. These two examples are designed to simulate the cases of Figs. 2.4 and 2.5 of Barnett and Radmore,[3] and qualitative agreement is found although the present time appears to be scaled from their plot labels. The regeneration of $p_s(t)$ is vividly displayed in Figs. 1 and 3 and is discussed below.

The exponential for these figures is

$$p_s(t) = exp(-\gamma t) . \tag{11}$$

A straight edge is used to find $\gamma = 0.26$ for Fig. 2. When a continuum of states is present, then Refs. 8 and 12 find, in the present notation,

$$\gamma = 2\pi \bar{V}^2/\varepsilon . \tag{12}$$

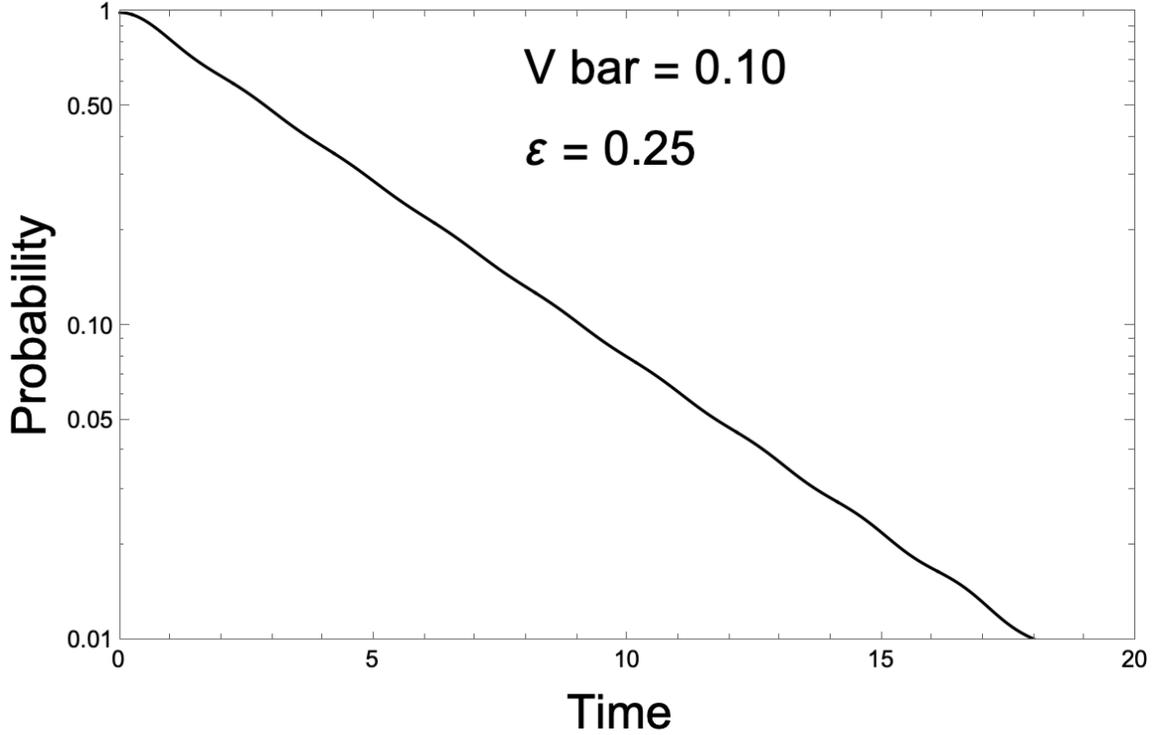

Fig. 2. The occupation of the initial state versus time for $\bar{V} = 0.10$ and $\varepsilon = 0.25$. This semi-logarithmic plot of $p_s(t)$ from Fig. 1 shows exponential decay although there are fine oscillations around the exponential.

This equals 0.251 for the case of Fig. 2. Similar results for the case of Fig. 3 yield 0.042 from a semi-logarithmic plot and 0.040 from Eq. (12). The agreement is reasonable enough that it suffices to assume that the present 25 states adequately represent a continuum.

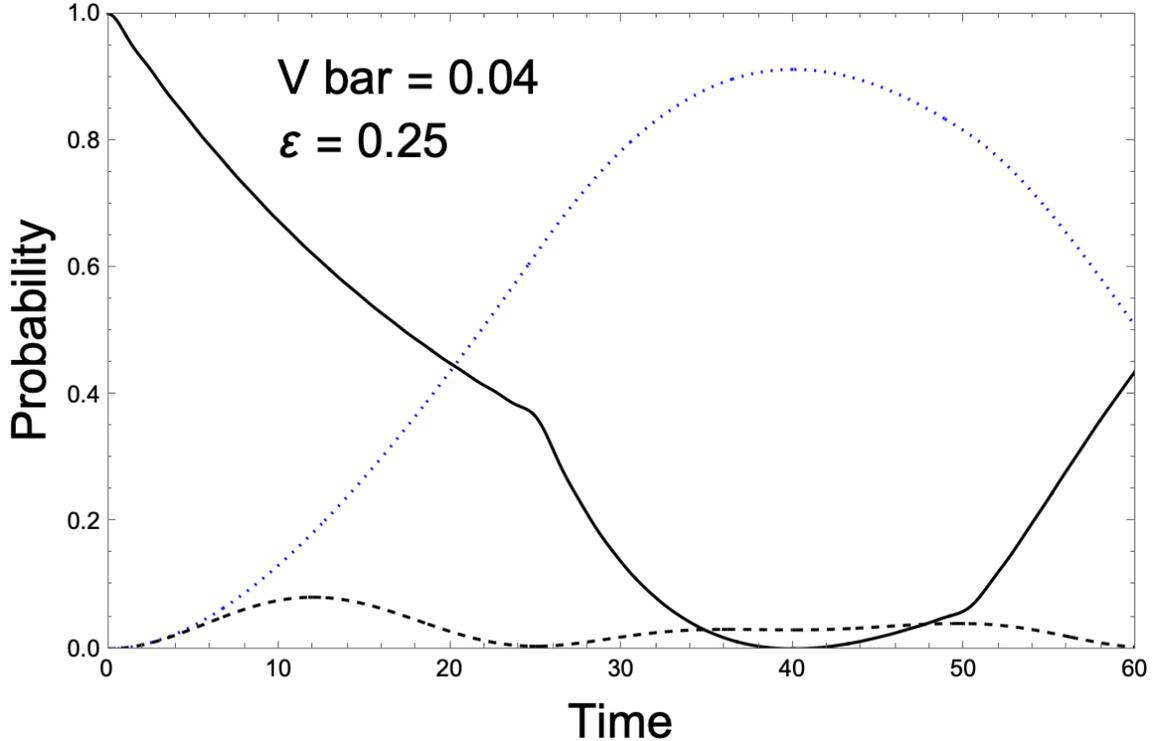

Fig. 3. The occupation of the initial state versus time for $\bar{V} = 0.04$ and $\varepsilon = 0.25$ with $n = 26$. The solid line is $p_s(t)$, the dotted line (blue) is the occupation probability for state $k = 0$, and the dashed line is the occupation probability for $k = 1$.

The numerical work continues with a study of how $p_s(t)$ changes when fewer than 26 states are used. This is followed by an exploration of the effects on $p_s(t)$ when the strength of the transition matrix element is varied with $n = 26$ states. The behavior of the $p_k(t)$ is also shown.

First, $n$ is varied with $\bar{V} = 0.10$ and $\varepsilon = 0.25$. This series starts with Fig. 1 for $n = 26$. This is followed by Fig. 4 with $n = 16$, which shows the results for $p_s(t), p_0(t)$, and $p_1(t)$ are quite similar to those for $n = 26$. The semi-logarithmic plot for $n = 16$ and $p_s(t)$, which is not shown, has more visible departures from linearity than Fig. 2. These trends continue for $n = 10$ and $n = 8$, which are also not shown. However, the semi-logarithmic plot for $n = 8$ appears in Fig. 5 and shows how the initial drop of $p_s(t)$ is getting rougher. Figure 6 has the semi-logarithmic plot for $n = 6$ and the initial drop of $p_s(t)$ is not exponential. Next, the results for $n = 4$ appear in Fig. 7. A comparison with Fig. 4 shows the growth of $p_0(t)$ and $p_1(t)$ as $n$ decreases. This is expected as there are fewer states to share the occupation probability of $1 - p_s(t)$. In addition, the second peak of $p_s(t)$ increases when $n$ drops.

Finally, $n = 2$ is not shown, but $p_s(t)$ and $p_0(t)$ swap total populations in the manner of a two-level system.[15] One might view this as the ultimate in regeneration or recurrence.

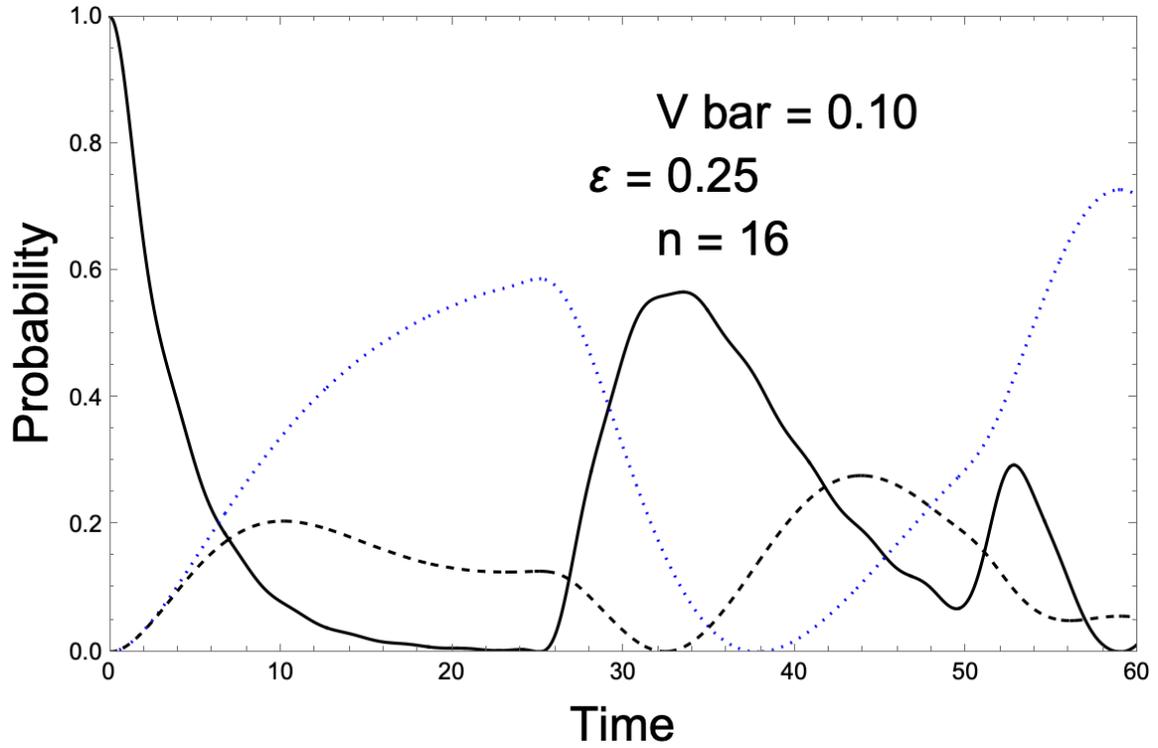

Fig. 4. The occupation of the initial state versus time for $\bar{V} = 0.10$ and $\varepsilon = 0.25$ with $n = 16$. The solid line is $p_s(t)$, the dotted line (blue) is the occupation probability for state $k = 0$, and the dashed line is the occupation probability for $k = 1$.

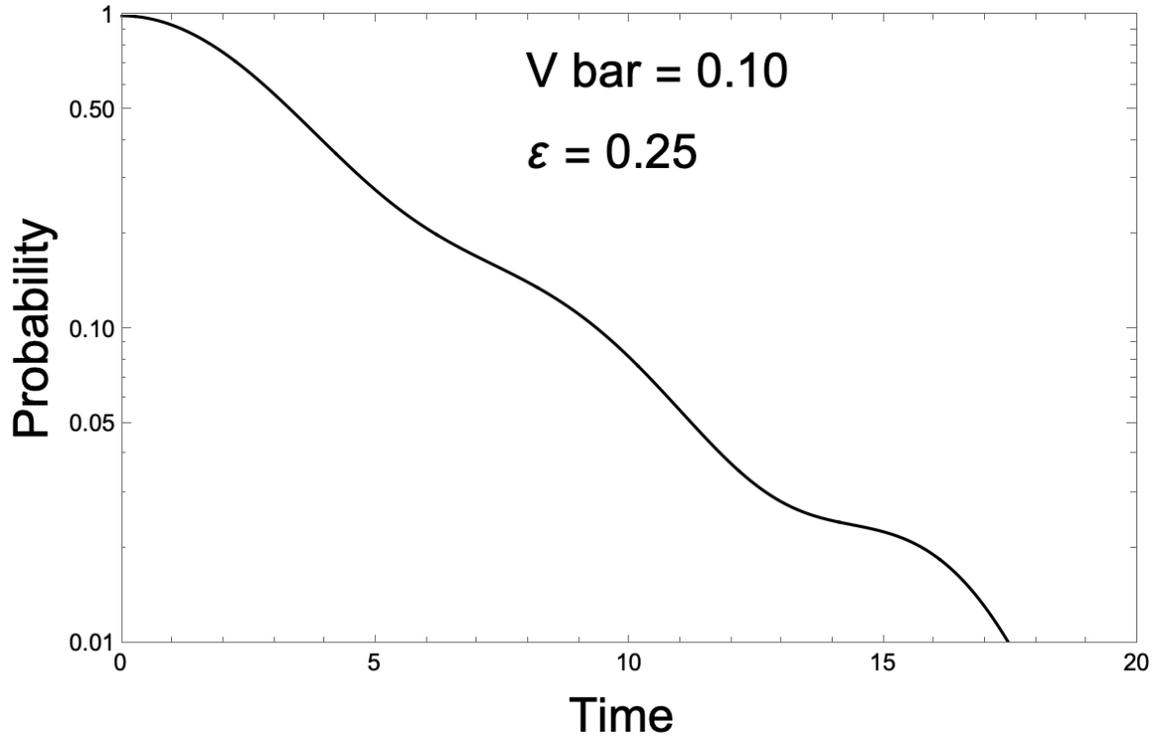

Fig. 5. The occupation of the initial state versus time for $\bar{V} = 0.10$ and $\varepsilon = 0.25$ with $n = 8$. This is a semi-logarithmic plot of $p_s(t)$ to show the oscillations around an initial trend of exponential decay.

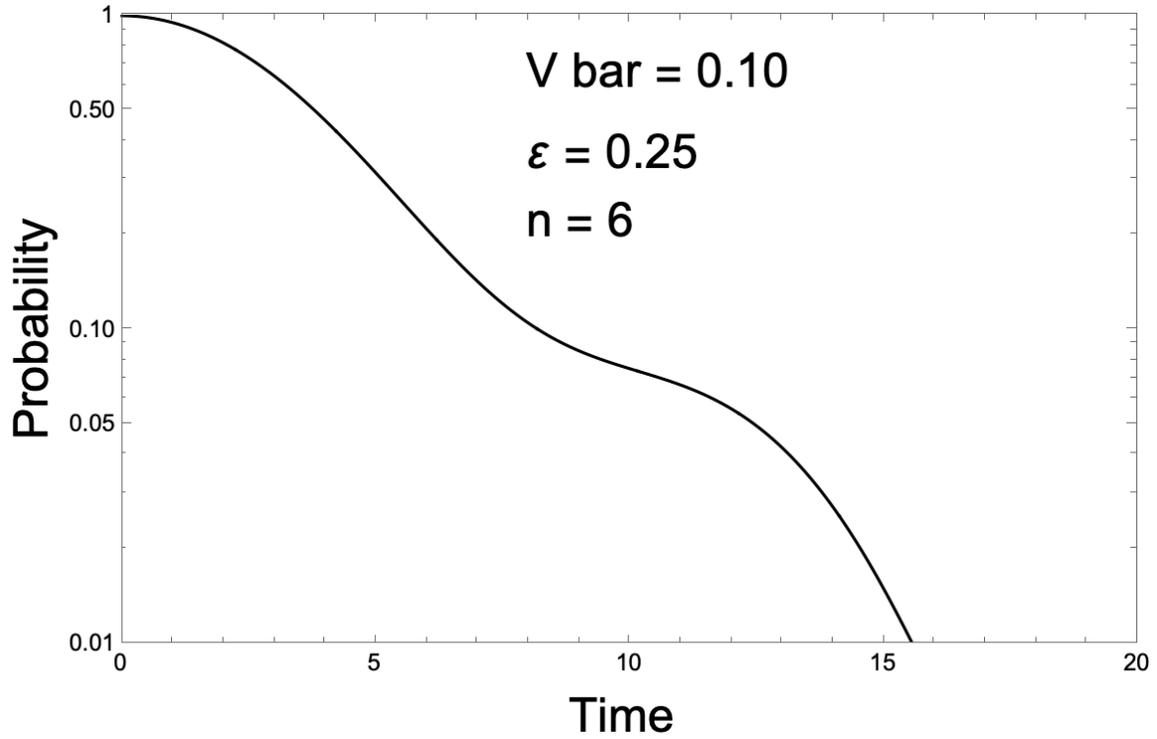

Fig. 6. The occupation of the initial state versus time for $\bar{V} = 0.10$ and $\varepsilon = 0.25$ with $n = 6$. This is a semi-logarithmic plot of $p_s(t)$ to show the departure from exponential decay.

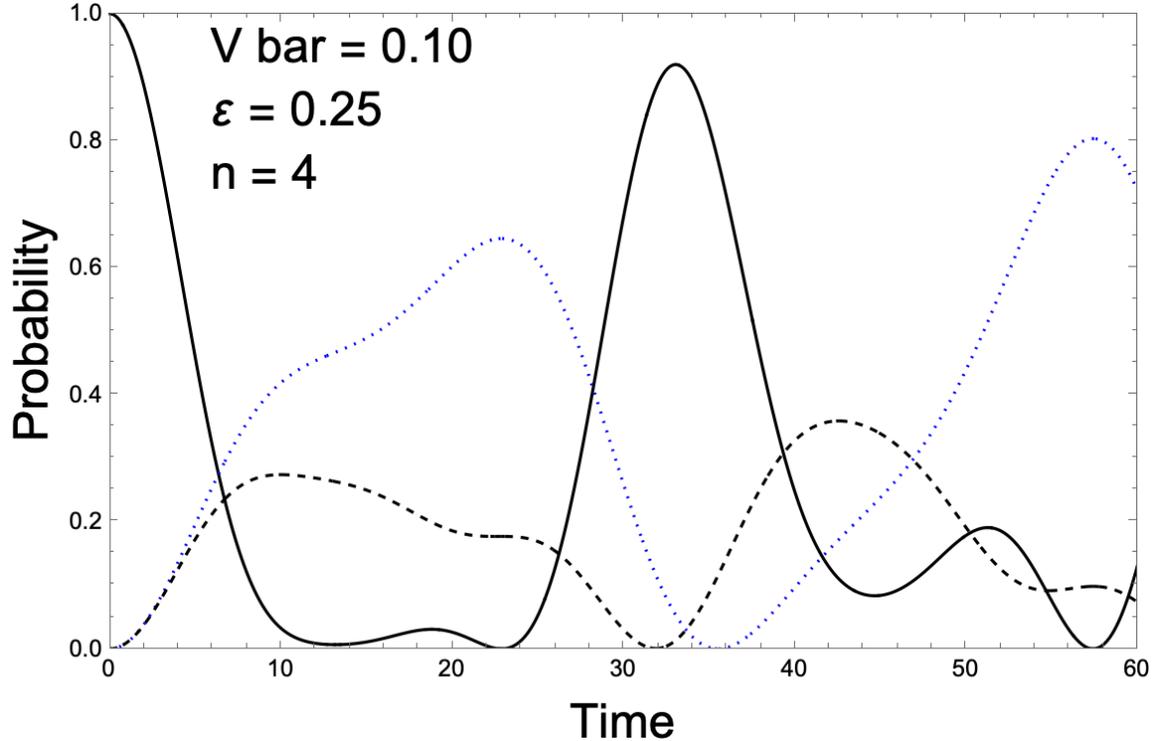

Fig. 7. The occupation of the initial state versus time for $\bar{V} = 0.10$ and $\varepsilon = 0.25$ with $n = 4$. The solid line is $p_s(t)$, the dotted line (blue) is the occupation probability for state $k = 0$, and the dashed line is the occupation probability for $k = 1$.

Next, $n$ is returned to 26 to explore the effects of varying the transition matrix element. The parameters are $\varepsilon = 0.10$ and $\bar{V} \leq 0.10$. The initial plot, Fig. 8, has $\bar{V} = 0.10$. The time is extended to 120 to catch the second peak. The transition matrix element is now the same size as the energy level separation, so numerous $k$-states are occupied. Figure 9 shows the initial decay is not exponential. The value of $\bar{V}$ is reduced to 0.075 for Fig. 10, which resembles Fig. 8 except that the second peak is lower and the $p_k(t)$ are decreasing faster with $k$. The semi-logarithmic plot for this case appears in Fig. 11 and an exponential decay is seen on average.

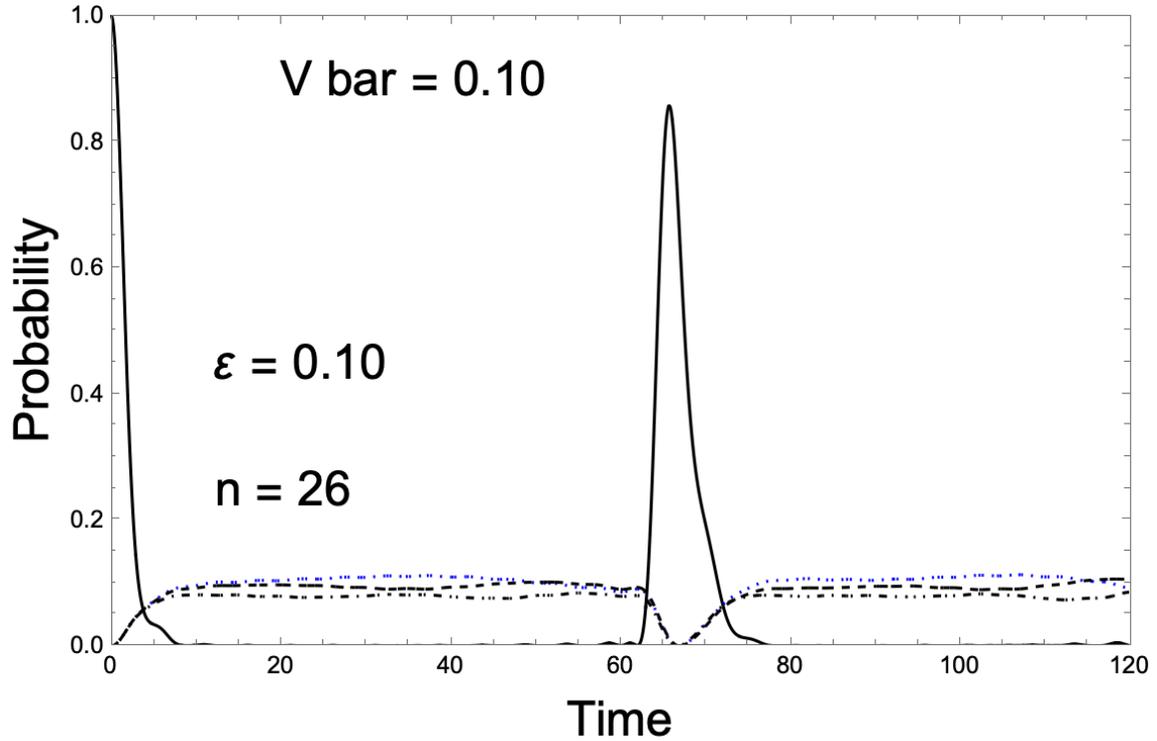

Fig. 8. The occupation of the initial state versus time for $\bar{V} = 0.10$ and $\varepsilon = 0.10$ with $n = 26$. The solid line is $p_s(t)$, the dotted line (blue) is the occupation probability for state $k = 0$, the dashed line is the occupation probability for $k = 1$, and the dash-dot line is the occupation probability for $k = 2$.

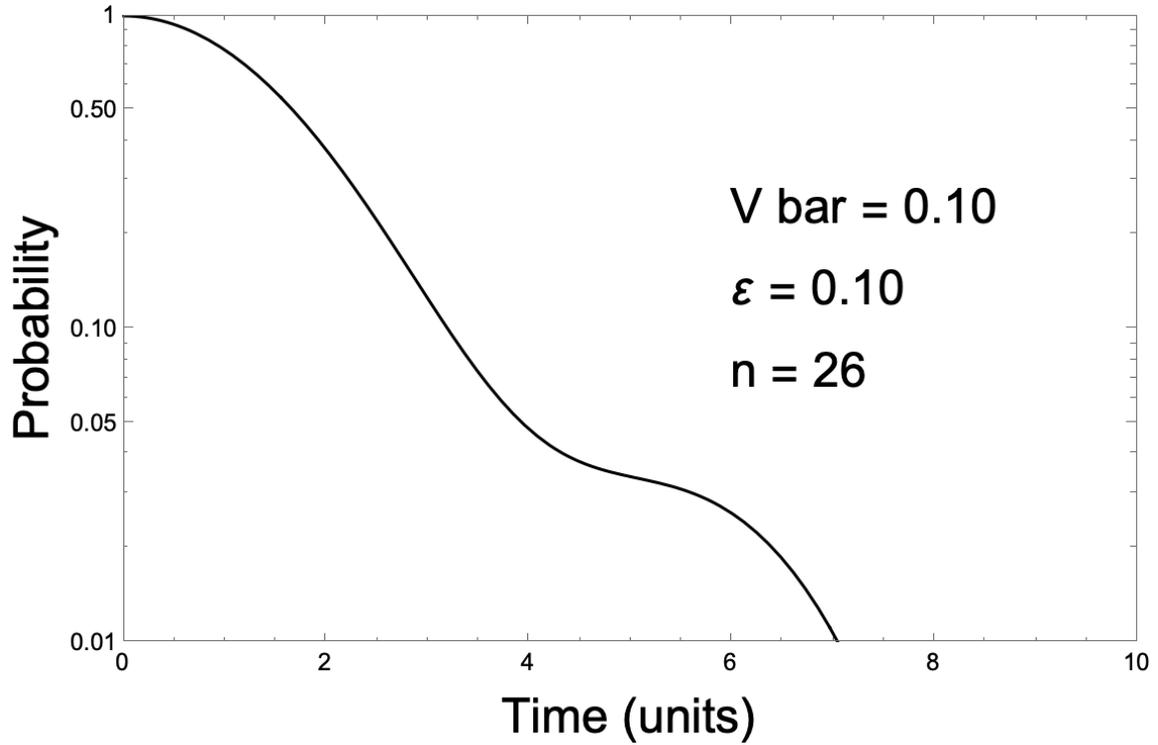

Fig. 9. The occupation of the initial state versus time for $\bar{V} = 0.10$ and $\varepsilon = 0.10$ with $n = 26$. This is a semi-logarithmic plot of $p_s(t)$ that shows the initial decay is not exponential.

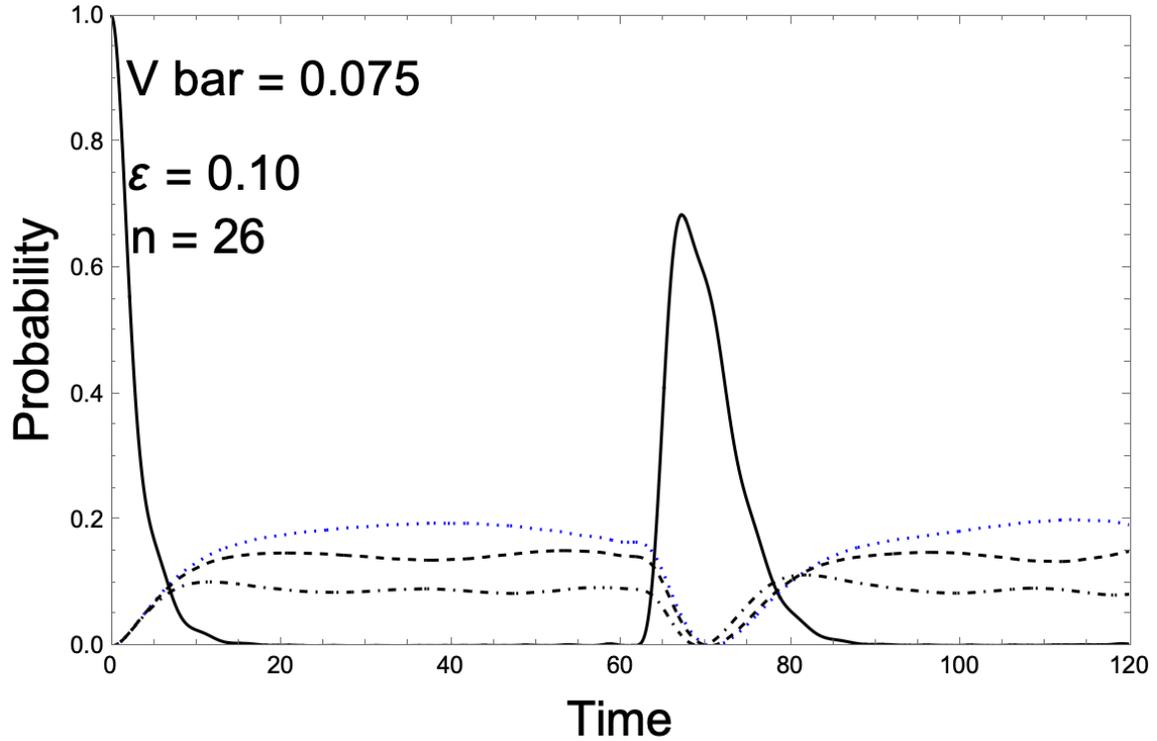

Fig. 10. The occupation of the initial state versus time for $\bar{V} = 0.075$ and $\varepsilon = 0.10$ with n = 26. The solid line is $p_s(t)$, the dotted line (blue) is the occupation probability for state $k = 0$, the dashed line is the occupation probability for $k = 1$, and the dash-dot line is the occupation probability for $k = 2$.

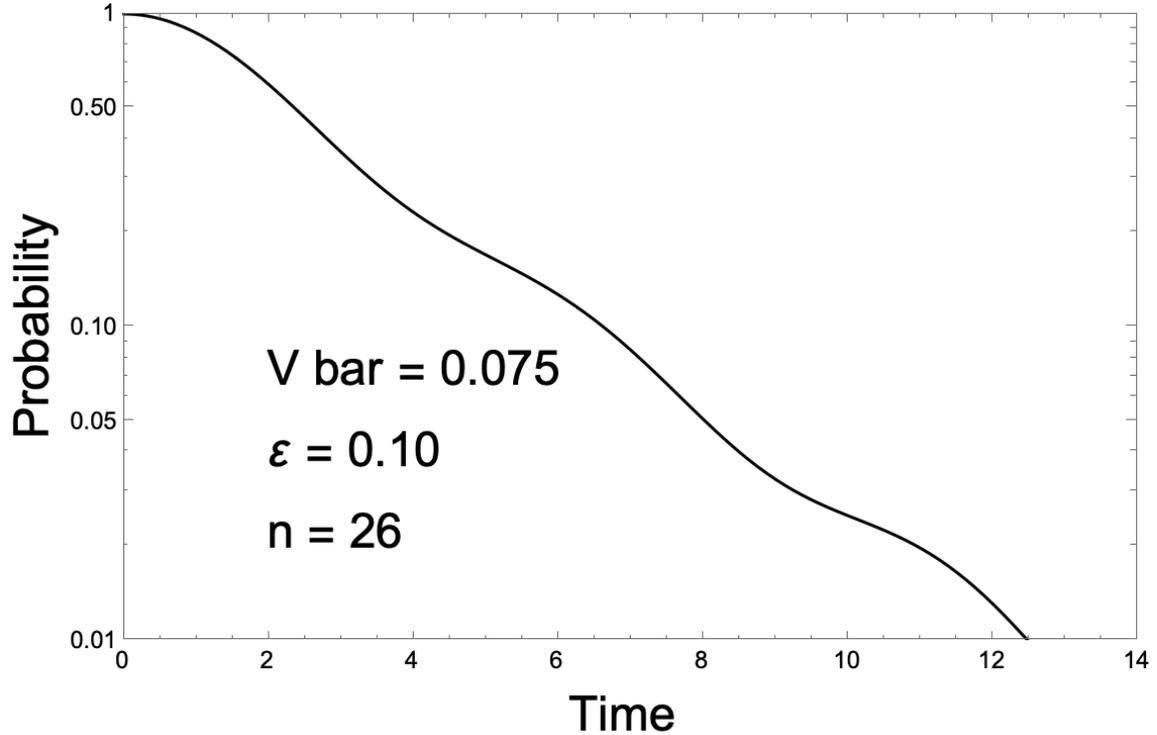

Fig. 11. The occupation of the initial state versus time for $\bar{V} = 0.075$ and $\varepsilon = 0.10$ with $n = 26$. This is a semi-logarithmic plot of $p_s(t)$ that shows the initial decay is roughly exponential.

These trends continue with a further decrease of $\bar{V}$. The time needed for the first decrease in $p_s(t)$ keeps increasing and the location of the second peak of $p_s(t)$ moves out in time. Further, the break in the initial drop of $p_s(t)$ becomes more apparent. These points are illustrated in Fig. 12 for $\bar{V} = 0.02$. In addition, the $p_k(t)$ are dominated by $p_0(t)$, which is expected when the strength of the transition matrix element decreases. The system then acts more like a two-level system with the states $s$ and $k = 0$. This coincides with the location of the first maximum of $p_0(t)$ moving to larger times. Figure 13 is the semi-logarithmic plot for $p_s(t)$ with $\bar{V} = 0.02$ and the break is clearly revealed for $t \approx 60$.

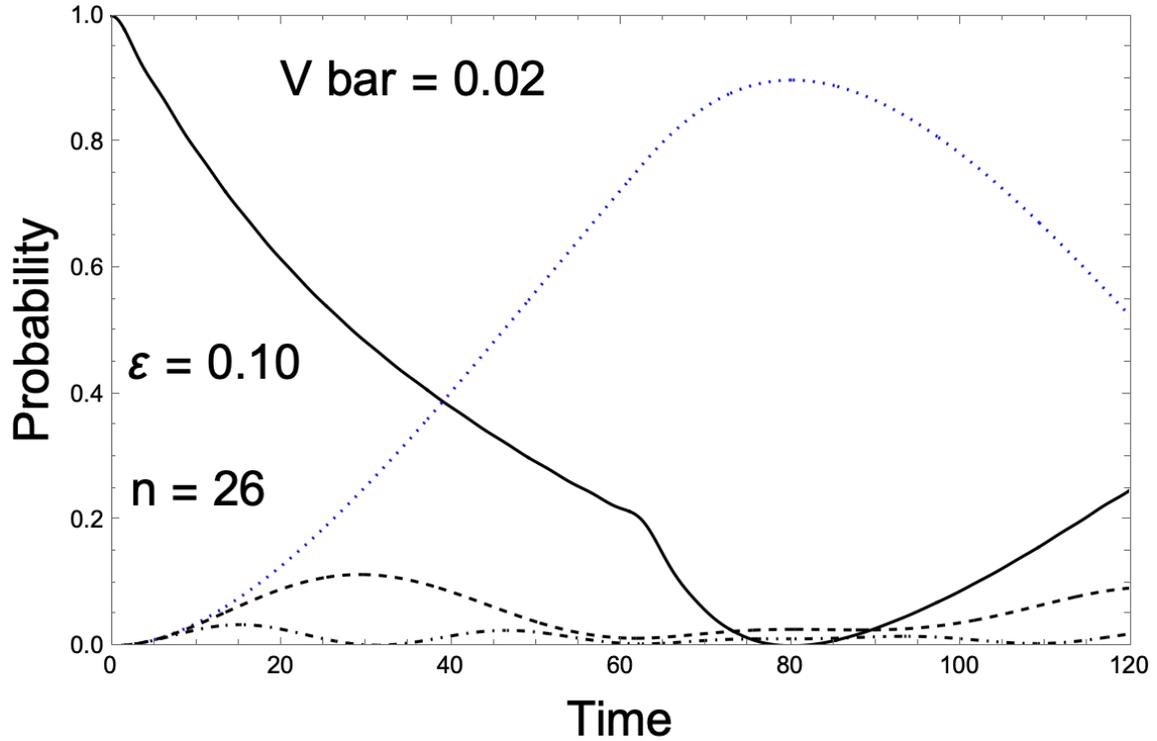

Fig. 12. The occupation of the initial state versus time for $\bar{V} = 0.02$ and $\varepsilon = 0.10$ with $n = 26$. The solid line is $p_s(t)$, the dotted line (blue) is the occupation probability for state $k = 0$, the dashed line is the occupation probability for $k = 1$, and the dash-dot line is the occupation probability for $k = 2$.

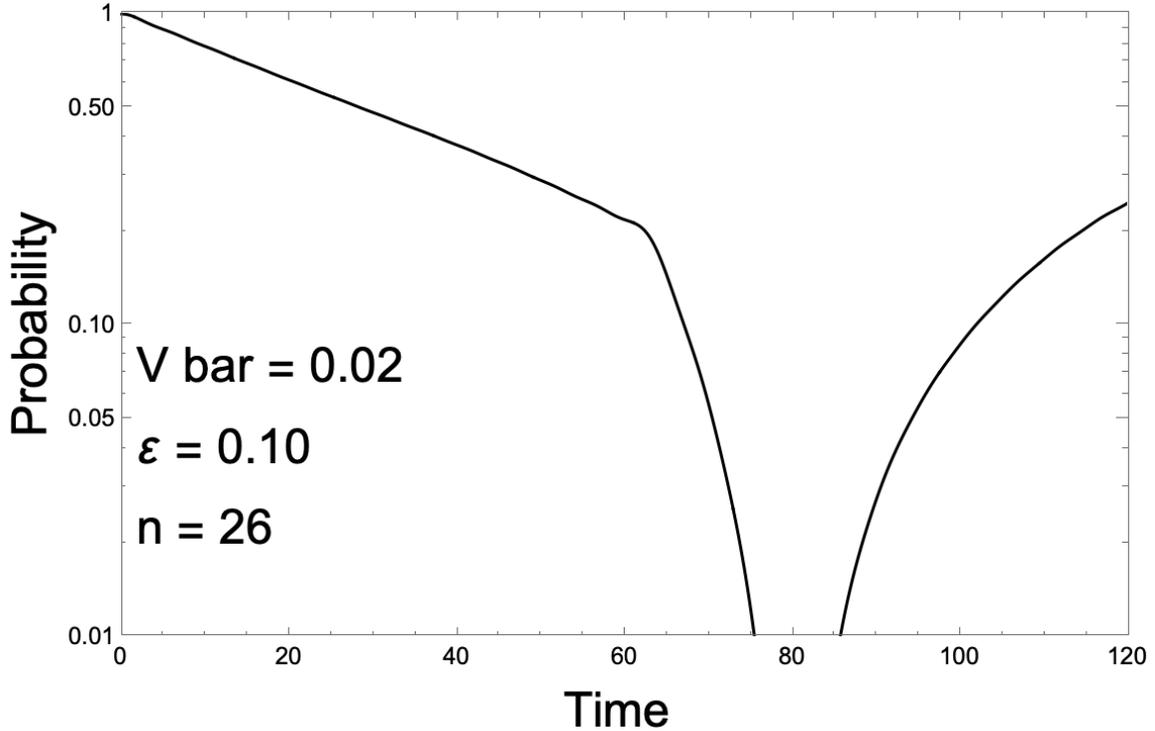

Fig. 13. The occupation of the initial state versus time for $\bar{V} = 0.02$ and $\varepsilon = 0.10$ with $n = 26$. This semi-logarithmic plot of $p_s(t)$ shows the initial decay is exponential before the break.

A sense of how the changes in time play out for weaker $\bar{V}$ is provided by Fig. 14 with $\bar{V} = 0.002$. Here, $p_1(t)$ and $p_2(t)$ shadow the horizontal axis, which extends to $t = 240$. The first peak in $p_0(t)$ occurs at still larger t.

It is worth considering how the $p_k(t)$ increase from their zero values at $t = 0$. The case studied has $\varepsilon = 0.10$ and $\bar{V} = 0.075$ with $n = 26$. Figure 10 shows that $p_0(t), p_1(t)$, and $p_2(t)$ all appear to rise at a similar time and they then peak at lower values as the time increases. The $p_k(t)$ for larger k peak even earlier in time and at lower levels before they start their regrowth. This behavior is illustrated in Fig. 15, which is a blow-up of the vertical axis with plots for $p_0(t), p_8(t)$, and $p_{12}(t)$. These details help to explain the rises and falls of $p_s(t)$. If one is checking $1 - p_s(t)$, one needs to note that $p_k(t) = -p_k(t)$ in the Bixon-Jortner model for $k > 0$, and the present calculations have verified this.

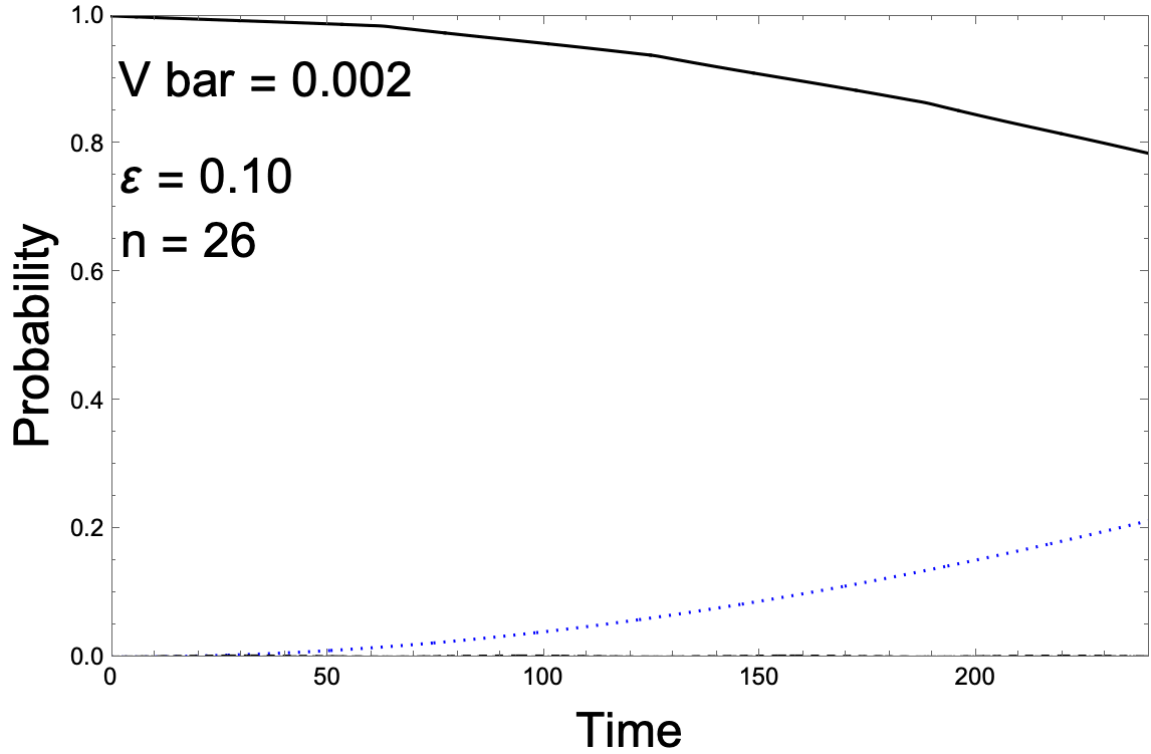

Fig. 14. The occupation of the initial state versus time for $\bar{V} = 0.002$ and $\varepsilon = 0.10$ with $n = 26$. The solid line is $p_s(t)$ and the dotted line (blue) is the occupation probability for state $k = 0$. The dashed line is the occupation probability for $k = 1$, and the dash-dot line is the occupation probability for $k = 2$. The latter two lines overlay the horizontal axis. The time goes to 240.

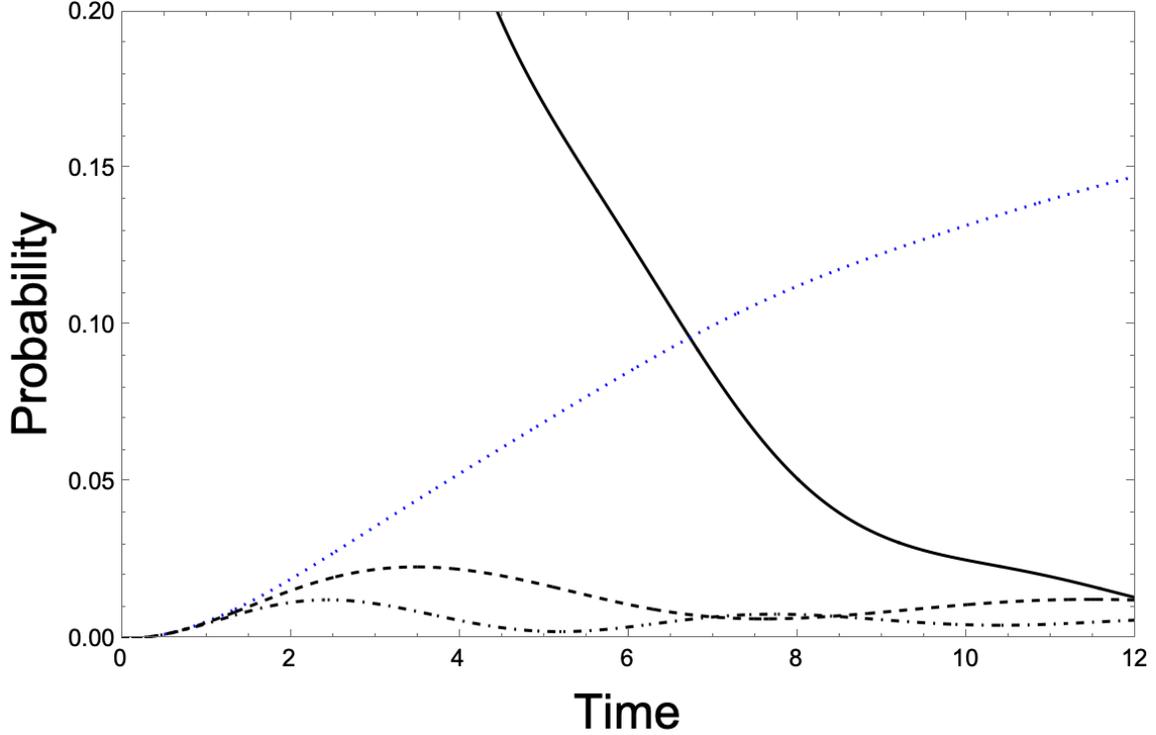

Fig. 15. The occupation of selected $k$-states versus time for $\bar{V} = 0.075$ and $\varepsilon = 0.10$ with $n = 26$. The solid line is $p_s(t)$, the dotted line (blue) is the occupation probability for state $k = 0$, the dashed line is the occupation probability for $k = 8$, and the dash-dot line is the occupation probability for $k = 12$. The vertical axis goes from 0 to 0.20.

This series with $\varepsilon = 0.10$ and $\bar{V} < 0.10$ shows $p_s(t)$ decays exponentially or approximately exponentially from just past $t = 0$. The slope of each decay is found with a straight edge and a semi-logarithmic plot. These slopes are compared with those from Eq. (12) in Table I.

| $\bar{V}$ | $\gamma$ from plot | $\gamma$ from Eq. (12) |
|---|---|---|
| 0.075 | 0.38 | 0.353 |
| 0.05 | 0.16 | 0.157 |
| 0.02 | 0.025 | 0.0251 |
| 0.01 | 0.0065 | 0.00628 |
| 0.002 | 0.00037 | 0.00025 |

Table I. Factor for the exponential decay of $p_s(t)$ – plot versus theory for a continuum of states.

The table shows good agreement except for $\bar{V} = 0.002$. The semi-logarithmic plot of $p_s(t)$ for this case displays a very slow change of slope with time and the extraction of $\gamma$ is from the first straight part of the decay. Its projection to $t = 240$ gives $p_s(t) = 0.915$.

Further discussion of the numerical results occurs in the next section.

## IV CONCLUSIONS

Section III illustrates how an initial state decays into a finite number of other states in the Bixon-Jortner model. The survival or occupation probability, $p_s(t)$, of this initial state decays, but often the decay is not exponential in time. As the time increases, the initial state is repopulated. This is in accord with the quantum recurrence theorem[4-7] that applies when the number of states is countable and not a continuum. Figure 4 shows how the other states such as those for $k = 0$ and $k = 1$, have populations that first grow when $p_s(t)$ decreases and then decrease when $p_s(t)$ increases. This behavior is also seen in Figs. 7, 8, 10, 12, and 15 and reflects the conservation of probability. In addition, when $V$ decreases with respect to $\varepsilon$, the system more strongly approximates a two-level system with the state $s$ and the state $k = 0$.

Some of the cases do exhibit a $p_s(t)$ with an exponential decay after the small-$t$ region. These are cases with the energy increment $\varepsilon > \bar{V}$. Even in these cases, $p_s(t)$ departs from the exponential as the time $t \to 0$ in agreement with arguments[4] based on the finiteness of $\langle s|\hat{H}|s \rangle$, as it is here. Basically, these arguments lead to the time derivative of $p_s(t)$ being zero at $t = 0$, while exponential decay yields a negative derivative there.

This paper has explored and implemented the Bixon-Jortner model for a finite set of states. The resulting survival probability, $p_s(t)$, of the initial state shows a time development in line with the expectations for solutions of Schrödinger's Equation. The key features are the presence of regeneration or recurrence and the frequent lack of a time interval with exponential decay for $p_s(t)$.

Further work needs to utilize more realistic transition matrix elements, and allow transitions between states. Winter[17] has provided one such model based on barrier penetration in one dimension. Further development of this model has taken place.[18,19] Whether one is able to find other models with some of these generalizations that are also "solvable" is a question for the future.

ACKNOWLEDGMENTS

I thank Quang Ho-Kim and Harvey S. Picker for helpful discussions. This manuscript is taken from an earlier version and incorporates many suggestions from an anonymous reviewer. I thank the reviewer.

**AUTHOR DECLARATION**
**Conflict of Interest**

The author has no conflicts to disclose.

**DATA AVAILABILITY**

The data that support the findings of this study are available within the article.